%% file: main.tex
\documentclass[conference]{IEEEtran}
\IEEEoverridecommandlockouts
\usepackage{times}

\usepackage[numbers]{natbib}
\usepackage{multicol}
\usepackage[bookmarks=true]{hyperref}

\usepackage[small]{caption}
\usepackage{graphicx}
\usepackage{amsmath}
\usepackage{amsthm}
\usepackage{booktabs}
\usepackage{algorithm2e}
\usepackage{balance}

\usepackage{xcolor}
\usepackage{subcaption}
\usepackage{amsmath,amssymb,amsfonts}
\usepackage{graphicx}
\usepackage{textcomp}
\usepackage{xcolor}
\def\BibTeX{{\rm B\kern-.05em{\sc i\kern-.025em b}\kern-.08em
    T\kern-.1667em\lower.7ex\hbox{E}\kern-.125emX}}

\usepackage{manyfoot}

\DeclareNewFootnote{A}
\DeclareNewFootnote{B}

\let\footnoteR\footnoteB
\let\footnote\footnoteA
    
\begin{document}

\title{Group Decision-Making in Robot Teleoperation:\\ Two Heads are Better Than One}


\newcommand{\an}[1]{\textcolor{teal}{(An:  #1)}}
\author{\IEEEauthorblockN{Duc-An Nguyen \IEEEauthorrefmark{1}}
\IEEEauthorblockA{\textit{Oxford Robotics Institute} \\
\textit{University of Oxford}\\
Oxford, United Kingdom \\
annguyen@robots.ox.ac.uk}
\and
\IEEEauthorblockN{Raunak Bhattacharyya \IEEEauthorrefmark{1}}
\IEEEauthorblockA{\textit{Yardi School of Artificial Intelligence} \\
\textit{Indian Institute of Technology Delhi}\\
New Delhi, India \\
raunakbh@iitd.ac.in}
\and
\IEEEauthorblockN{Clara Colombatto}
\IEEEauthorblockA{\textit{Department of Psychology} \\
\textit{University of Waterloo}\\
Ontario, Canada \\
clara.colombatto@uwaterloo.ca}
\and
\IEEEauthorblockN{Steve Fleming}
\IEEEauthorblockA{\textit{Department of Experimental Psychology} \\
\textit{University College London}\\
London, United Kingdom \\
stephen.fleming@ucl.ac.uk}
\and
\IEEEauthorblockN{Ingmar Posner}
\IEEEauthorblockA{\textit{Oxford Robotics Institute} \\
\textit{University of Oxford}\\
Oxford, United Kingdom \\
ingmar@robots.ox.ac.uk}
\and
\IEEEauthorblockN{Nick Hawes}
\IEEEauthorblockA{\textit{Oxford Robotics Institute} \\
\textit{University of Oxford}\\
Oxford, United Kingdom \\
nickh@robots.ox.ac.uk
}}



%


\maketitle

\begingroup\renewcommand\thefootnote{\IEEEauthorrefmark{1}}
\footnotetext{Both authors contributed equally to this paper.}
\input{sections/intro/abstract}
\input{sections/intro/intro}

\input{sections/relatedwork/relatedwork}

\input{sections/experiment/experiment}

\input{sections/results/results}

\input{sections/discussion/discussion}

\bibliographystyle{plainnat}
\balance
\bibliography{Biblio}
\end{document}

%% file: sections/intro/abstract.tex
\begin{abstract}
Operators working with robots in safety-critical domains have to make decisions under uncertainty, which remains a challenging problem for a single human operator. An open question is whether two human operators can make better decisions jointly, as compared to a single operator alone. While prior work has shown that two heads are better than one, such studies have been mostly limited to static and passive tasks. We investigate joint decision-making in a dynamic task involving humans teleoperating robots. We conduct a human-subject experiment with $N=100$ participants where each participant performed a navigation task with two mobiles robots in simulation. We find that joint decision-making through confidence sharing improves dyad performance beyond the better-performing individual ($p<0.0001$). Further, we find that the extent of this benefit is regulated both by the skill level of each individual, as well as how well-calibrated their confidence estimates are. Finally, we present findings on characterising the human-human dyad's confidence calibration based on the individuals constituting the dyad. Our findings demonstrate for the first time that two heads are better than one, even on a spatiotemporal task which includes active operator control of robots.
\end{abstract}

%% file: sections/intro/intro.tex
\section{INTRODUCTION}
\input{sections/experiment/simulator}

Human operators are increasingly collaborating with robots via teleoperation in domains such as inspection~\cite{hawes2017strands,budd2023bayesian, chiou2016experimental, chiou2021mixed, corredor2016decision, shen2004collaborative}, nuclear decommissioning~\cite{nagatani2013emergency,chiou2022robot}, and search and rescue~\cite{casper2003human,dole2015robots, li2015role, music2017robot}. In these complex environments, operators are often faced with the decision of choosing which robot or robot controller to operate. For instance, an operator may need to select a robot from a fleet to assist in case of failure~\cite{ji2022traversing} or choose which robot to teleoperate~\cite{lee2016development, hoque2023fleet,nguyen2020model} and the level of autonomy for it to operate at~\cite{nam2019models,milliken2017modeling, nguyen2020model,nguyen2020modeling} during missions under time pressure and uncertainty. Decisions like these are complicated by communication latencies, incomplete information, and the dynamic nature of the environment, making the choice of which robot to assist or teleoperate a challenging and cognitively demanding task. 
In such scenarios, single operators can become overwhelmed, especially when required to make decisions under stress ~\cite{chen2012supervisory, liu2024affects, reed2008physical,reinoso2008mechanisms,nguyen2020modeling}. Individual human operators have limitations such as cognitive overload and biases. This raises an important question: \textbf{Could two decision-makers working together make better decisions than one individual acting alone?} Collaborative decision-making may provide a solution to these limitations. While each operator has their own weaknesses, humans often possess complementary skills that, when combined, can offset each other's shortcomings. Several studies in human-human interaction (HHI) shows that collaboration can lead to better decision outcomes~\cite{Harada_2021,Koriat_2015, Mahmoodi_Bang2013,Pescetelli_Rees_Bahrami_2016,Rouault_Fleming_2018}, and there is a strong case for transferring HHI findings into human-robot interaction (HRI)~\cite{trafton2005enabling,lockerd2004tutelage,hoffman2007cost,sakita2004flexible,Li_Zhou_2022}.

In the aforementioned domains, many teleoperation scenarios already involve multiple humans operating a single robot~\cite{Ji_Campbell_2022,Dahiya_Smith_2021,Cai_Smith_2022,han2023crossing}. The ratio of people to robots directly affects the human-robot interaction in such systems. For example, in the mobile robot search and rescue operations using telerobots, the operator-to-robot ratio is commonly 2 to 1 or higher \cite{murphy2004human, scholtz2004evaluation}. In control rooms for Unmanned Aerial Vehicle missions, several operators are needed to operate a single drone. In these scenarios, specific roles are often assigned to different team members, such as a "pilot" responsible for navigation and control, and a sensor/payload operator managing cameras and other equipment \cite{drury2006decomposition}. In the multi-UAV control setting, \citeauthor{hughes2008human} et al. showed that combining multiple human operators with partially autonomous robots can create more robust and effective systems than either working alone. In these multi-operator, single-robot scenarios, team dynamics~\cite{burke2008toward, 8673306, hedayati2018improving}, communication~\cite{hinds2004whose, 7451744} and coordination strategies~\cite{aronson2024intentional,elbeleidy2022practical} are all critical to successful team performance.

In HHI collaborative cognition, a line of research on human-human dyadic joint decision-making has shown that teams can outperform individuals under specific conditions, particularly when the joint decisions are guided by the Maximum Confidence Slating (MCS) approach~\cite{Bahrami2012a,Bahrami2012b,Bang_Bahrami_2014,Martino_Fleming_Garrett_Dolan_2013,Fleming_Daw_2017}. This approach prioritizes the decision in which the participant has the \textit{higher confidence}, assuming that each individual can monitor their own performance and can communicate their confidence accurately~\cite{bahrami2010optimally,koriat2012two}. While MCS has been shown to improve performance in tasks such as visual perception or knowledge-based tasks, its application to active, dynamic decision-making in robotic control tasks has not been adequately studied. We address this gap by investigating the performance of MCS in a spatio-temporal task, where human operators control robots in a simulated environment. Unlike previous research, which focused on passive information processing, our study involves humans actively gathering and interpreting information to make real-time decisions.

Specifically, in our user study, we conducted experiments with 100 participants, each controlling two different robots in an online simulation. The environments and control latencies of the robots were altered, with one robot exhibiting a more favorable delay. Participants had to choose which robot was better for the task, based on the robot’s responsiveness and their own confidence in the choice. We then applied the MCS approach to compare the accuracy of joint decision-making of dyads being virtually paired with individual performance, using an \textit{accuracy gain}\footnoteR{Accuracy gain is defined as the accuracy increase/decrease difference between the accuracy (\%) of decision made by the MCS-based joint decision-making agent as compared with the higher performing individual in the dyad} metric.

To the best of our knowledge, this is the first study to apply MCS in a scenario where humans must actively control robots rather than passively receiving information.
Our findings reveal the following key insights
into the effectiveness of the MCS-based decision-making process in joint decision-making within robotics teleoperation domains:

\begin{enumerate}
    \item \textbf{Impact of Confidence on Accuracy:} Joint decisions determined by high confidence via MCS were more accurate than those made by the highest-performing individual in the dyad. Conversely, joint decisions based on low-confidence inputs were less reliable than random choices, emphasizing the need to avoid low-confidence responses.

    \item \textbf{Effect of Performance Discrepancy:} This is the first time we have observed that larger performance gaps between dyad members result in smaller accuracy gains from MCS. Pairing participants with similar performance levels yielded significantly higher accuracy, highlighting that MCS is more effective when participants have comparable competence.

    \item \textbf{Influence of Confidence Calibration:} Confidence calibration \cite{fleming2014measure} significantly impacted MCS accuracy gains. For individuals with above-average confidence calibration, MCS-based accuracy gains were stable regardless of calibration differences. For those with below-average calibration, pairing similarly calibrated individuals led to negative accuracy gains, while teaming individuals with diverse calibration levels improved the accuracy of the dyad. This marks the first time such analyses have been conducted on joint decision-making in dynamic tasks.

    \item \textbf{Correlation with Dyadic Confidence Calibration:} Accuracy gains from MCS were positively correlated with dyadic confidence calibration. Higher overall confidence calibration of the dyad led to better decision accuracy, further underscoring the importance of confidence calibration in joint decision-making.
\end{enumerate}


%% file: sections/experiment/simulator.tex
\begin{figure}
    \centering
    \includegraphics[width=0.9\columnwidth]{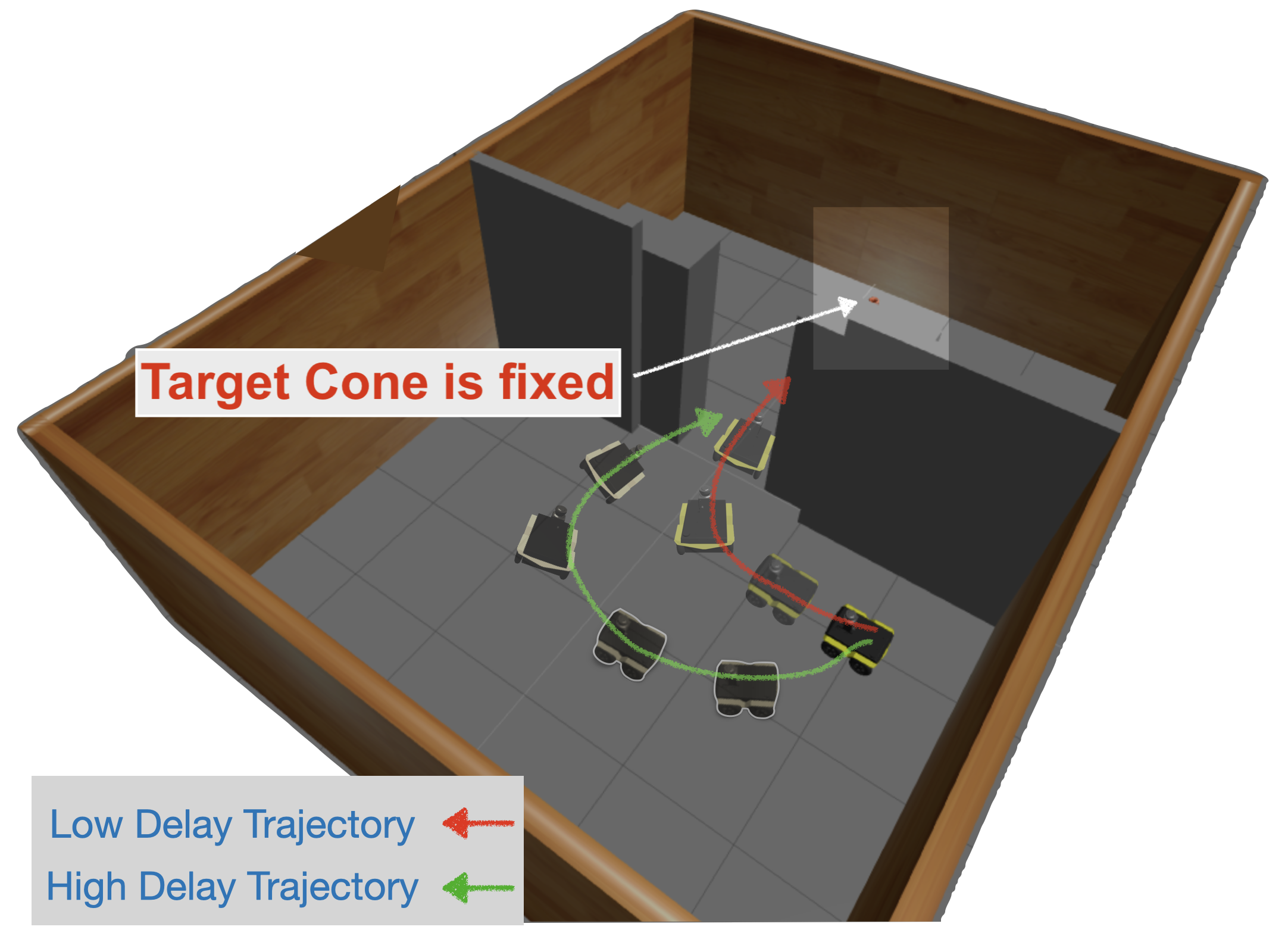}
    \caption{Web based robot navigation simulator. The red and green curves represent trajectories driven by a participant from our user study who was experiencing low and high delay in control of the mobile robot.}
    \label{fig:simulator}
\end{figure}

%% file: sections/relatedwork/relatedwork.tex
\section{RELATED WORK}

\input{sections/relatedwork/koriat}

%% file: sections/relatedwork/koriat.tex
\subsection{Multi-operator Teleoperation}

Multiple Operator Single Robot (MOSR) systems have emerged as a promising approach, particularly in the field of semiautonomous teleoperation \cite{ song2022costume, 9473689, 8534835,8542482, 6249573,6281254}. \citeauthor{reed2008physical} highlighted the potential benefits of MOSR systems via the studies of haptic human-human interaction in joint object manipulation tasks. It was shown that task performance of two humans solving a haptic task collaboratively is higher than that of a single operator performing the same task. The authors suggests that adding an additional human operator to a classical teleoperation scheme could have a positive effect on task performance \cite{monferrer2002}. 

Another example is the distributed teleoperation system developed by \citeauthor{goldberg2002collaborative}, where multiple users, each at a different location, simultaneously controlled an industrial robot arm over the internet. Their client-server system enabled multiple users to share control of a single robot, demonstrating how collaboration among users can enhance performance. Importantly, this research showed that human groups outperformed individuals when facing noisy environments in MOSR systems \cite{goldberg2000collaborative}. 

\subsection{Joint Decision-Making in Human Dyads}
\begin{figure*}[h]
\begin{subfigure}{0.33\textwidth}
    \includegraphics[width=1.05\textwidth]{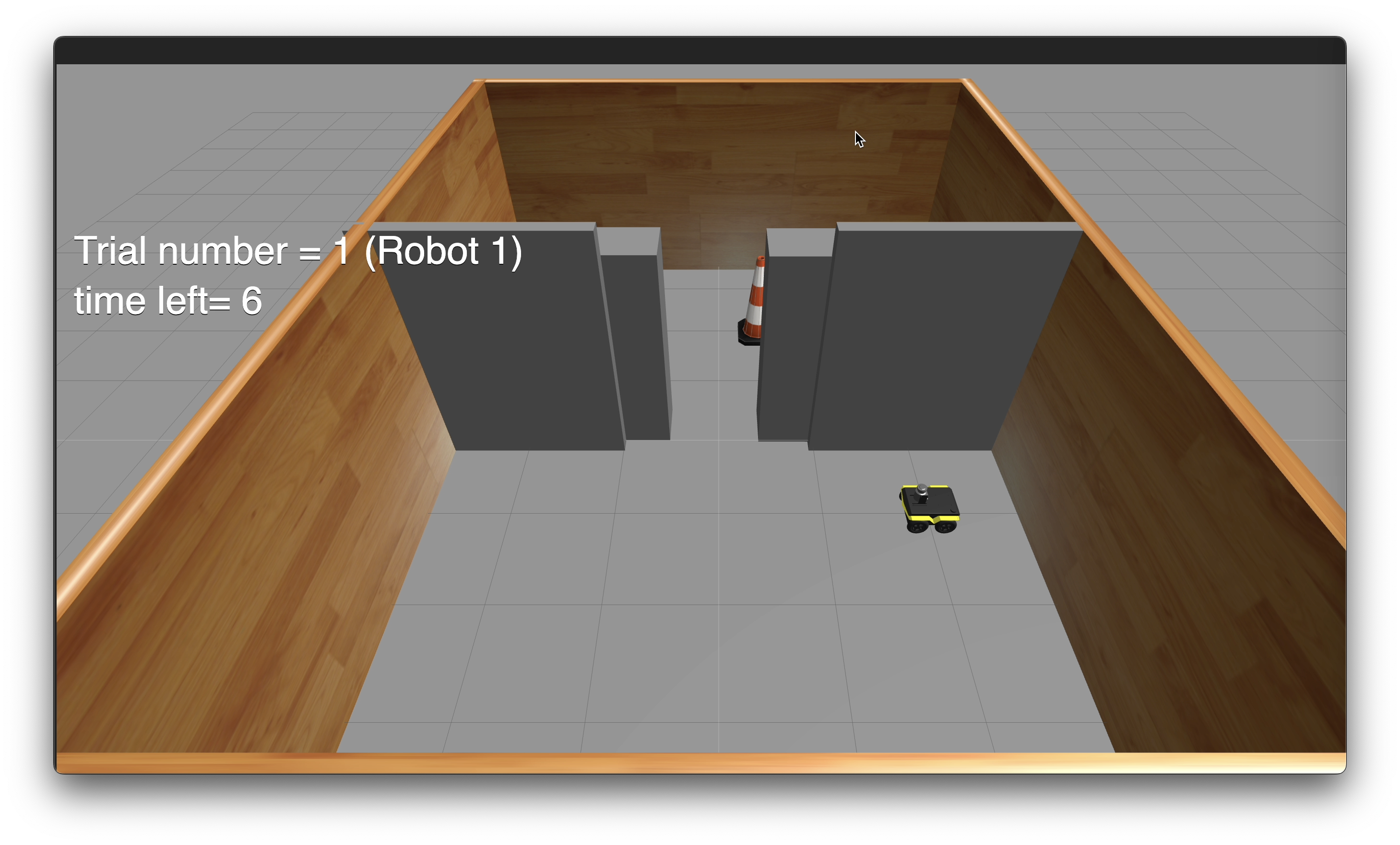}
    \caption{}
    \label{fig:procedure-2}
\end{subfigure}%
\begin{subfigure}{0.33\textwidth}
    \includegraphics[width=1.05\textwidth]{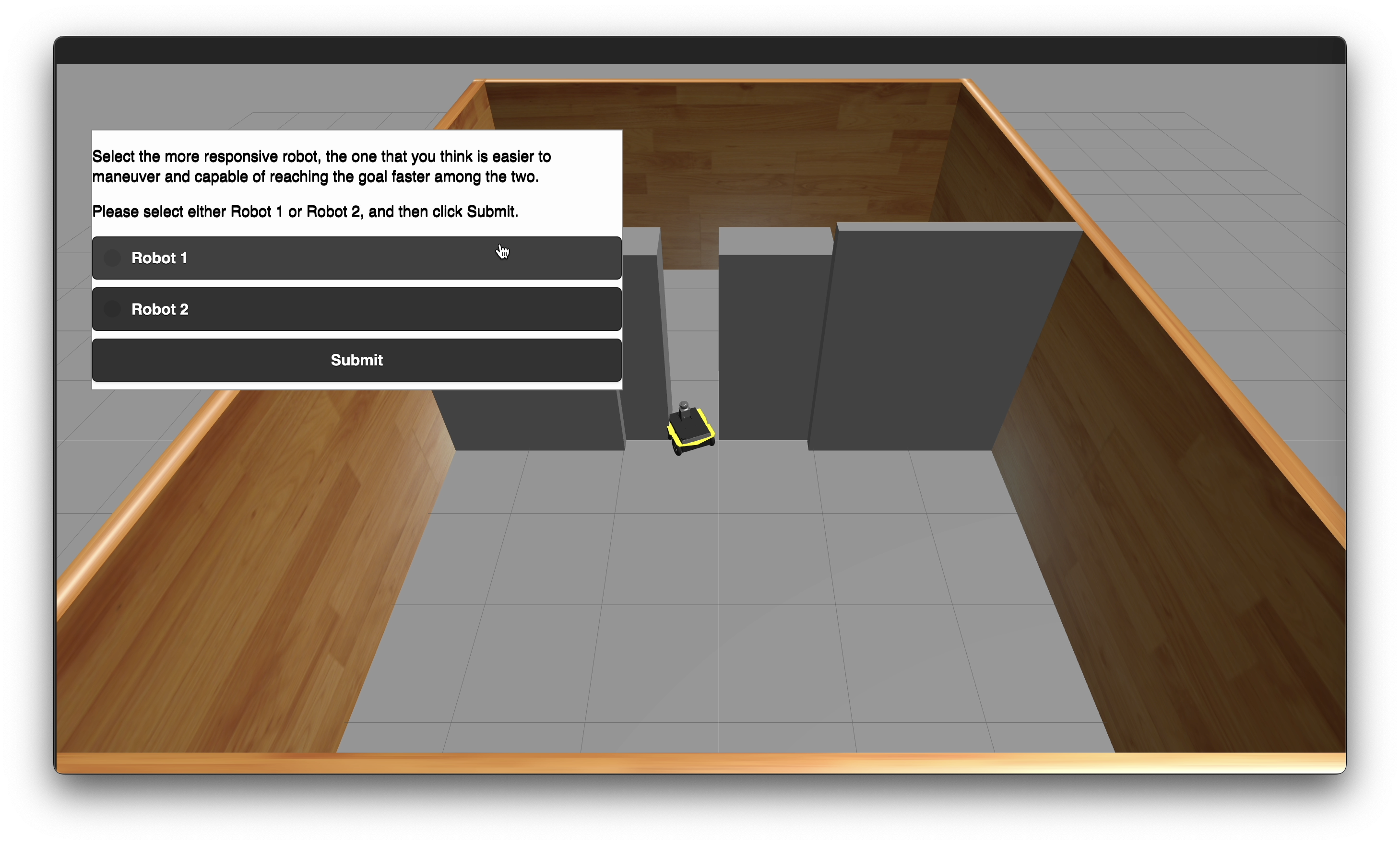}
    \caption{}
    \label{fig:procedure-4}
\end{subfigure}%
\begin{subfigure}{0.33\textwidth}
    \includegraphics[width=1.05\textwidth]{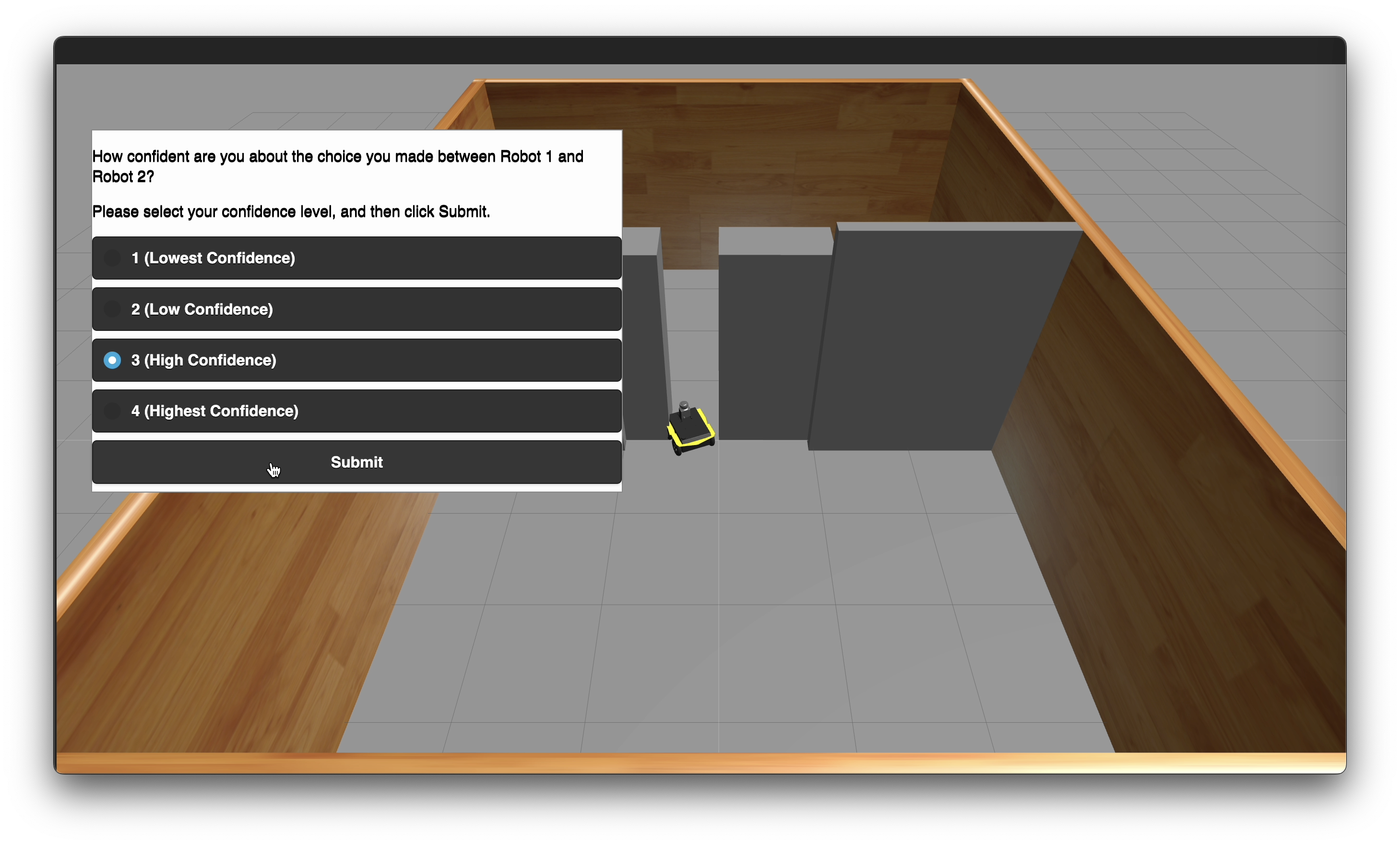}
    \caption{}
    \label{fig:procedure-5}
    \end{subfigure}

\caption{Screenshots showing driving trials scenario and question queries (\ref{fig:procedure-2}: environment setup; \ref{fig:procedure-4}: robot choices, \ref{fig:procedure-5}: user confidence levels) for each participant. There are total of 100 trials per participant.}
\label{fig:procedure}
\end{figure*}

Joint decision-making within human dyads has been extensively studied in the context of visual perception \cite{bahrami2010optimally,Bahrami2012a,Bahrami2012b,Bang_Aitchison,ma-pre-24,Martino_Fleming_Garrett_Dolan_2013}. \citeauthor{bahrami2010optimally} conducted experiments where participants, working in dyads, engaged in a two-alternative forced-choice task, deciding which of two briefly presented stimuli contained an oddball target. Participants initially made individual decisions before sharing their conclusions. In cases of disagreement, they discussed the matter until reaching a joint decision. The results confirmed that, given equal visual sensitivity, two individuals were indeed better than one, provided they could communicate freely. The authors suggested that this "two-heads-are-better-than-one" (2HABT1) effect depends on each participant's ability to monitor their own performance accuracy and accurately communicate their confidence level.

Further studies by \citeauthor{koriat2012two} demonstrated that the advantage of joint decision-making remains even when participants are unable to communicate directly~\cite{Bang_Bahrami_2014,massoni2017optimal,rouault2018psychiatric}. It is shown that decisions guided by the confidence heuristic—where the most confident choice is selected—can be just as accurate as decisions reached through direct interaction, especially for individuals with comparable reliability. This highlights the potential of confidence as a powerful tool in improving joint decision-making within dyads. Over time, the accuracy of collective decisions becomes indistinguishable between conditions with and without feedback. Although the learning process is slower without feedback, the eventual collective benefit matches that of situations where feedback is provided \cite{Bang_Bahrami_2014}. 
In line with these discoveries, studies in human-human joint decision-making on varied tasks such as threat detection by observing video feeds~\cite{2hbt1_threatdetection}, detecting fake news~\cite{2hbt1_fakenewsdetection, pham2023pre} deciding rank ordering between items on a survival situation task~\cite{2hbt1_nasasurvival}, and breast and skin cancer diagnosis~\cite{2hbt1_cancerdiagnosis}
have also shown that higher confidence decision selection leads to higher accuracy.

Building on the body of related work described above, we identify a gap in the research regarding dyadic joint decision-making in teleoperated robotic tasks, and formulate the following key research questions to guide our investigation into how humans can make better joint decisions through MCS in teleoperational robotics:
\par
\textbf{RQ1:} How does individual confidence affect the accuracy of joint decisions in dyadic settings, and what are the benefits of joint decision-making using MCS?
\vspace{2pt}
\par

\par
\textbf{RQ2:} How do performance discrepancies between dyad members affect the accuracy gains of MCS-based joint decisions?\vspace{2pt}
\par

\par
\textbf{RQ3:} How does the confidence calibration of participants influence the accuracy gains of MCS-based joint decisions?\vspace{2pt}
\par

\par
\textbf{RQ4:} What is the relationship between dyadic confidence calibration
and the accuracy of dyadic MCS-based joint decisions?\vspace{2pt}
\par

%% file: sections/experiment/experiment.tex
\section{METHODOLOGY}
We conducted an online study, approved by the University of Oxford Research Ethics Committee, to investigate how humans make joint decisions when selecting robot controllers.
\subsection{Experiment Setup}

The experiment tasked participants with navigating two visually identical simulated robots through a narrow gap using an interface shown in Figure~\ref{fig:simulator}. The environment consisted of 24 different conditions, combining 6 possible initial robot poses with 4 doorway configurations. A fixed goal location was marked by a traffic cone, with the initial robot pose and door configuration randomly sampled for each trial.
Participants were recruited via the Prolific Platform, which provided an overview of the study and ethics approval information. Those who agreed to participate and were at least 18 years old were shown a detailed video explaining the interface components and demonstrating example trials. Before the main experiment, participants completed 5 practice trials to familiarize themselves with the keyboard controls.

\subsection{Procedure}
\label{procedure}

The main experiment consisted of 100 trials. In each trial, participants controlled each robot for 6 seconds in a random order. This duration was determined through a separate pilot study as the minimum time needed for decision-making. To create a perceptible difference between the robots, we injected time delays into their controls. Randomly, one was assigned a fixed 50 ms delay, and the other a variable delay (70–150 ms).
We implemented a two-down one-up staircase\footnoteR{The two-down one-up staircase is a procedure where the task was made harder after 2 consecutive successful choices by decreasing the delay by one level step, and the task was made easier after 1 failure by increasing the delay by one level step.} procedure to adjust task difficulty.
This procedure ensured that the task was not so easy as to enable ceiling performance and not so hard to make participant performance close to random \cite{levitt1971transformed}. This precluded participants from being overly confident or underconfident on the task.
During each trial, participants controlled each robot sequentially, with a brief pause, and selected the robot with the lower delay, rating their confidence on a four-point Likert scale (1:lowest, 4:highest).
This experimental design allowed us to assess both decision-making accuracy and confidence calibration in human-robot interaction tasks under varying degrees of difficulty.
Figure~\ref{fig:procedure} depicts the trial procedure.

The data were later used to pair individuals into dyads virtually. 
For example, Participant 1 encountered 20 trials with the delay-pair condition: 50 ms delay for Robot 1 and 70 ms delay for Robot 2. Participant 2 encountered 15 trials with the same delay condition. Therefore, we could pair up to a total of $\frac{20 \times 15}{2} = 150$ trials, where Participants 1 and 2 were able to make virtually joint decisions.

\subsection{Data Collection}

To ensure data quality, we implemented exclusion criteria based on task performance and engagement. Participants with accuracy below 65\% or those who gave the same confidence rating for over 95 out of 100 trials were excluded. This threshold helped retain engaged participants while excluding those who were inattentive or ineffective. The exclusion criteria and performance metrics are consistent with standard procedures in human self-confidence research \cite{rouault2019forming,bhattacharyya2024towards, anil2023towards}, ensuring our results are comparable with existing literature in the field.
After applying these criteria, our final analysis included 80 participants (47 males, 33 females), with a mean age of 37 years (SD = 11) and an average experiment duration of 1.2 hours (SD = 0.25). Participants were compensated an average of USD 7.6.

\subsection{Metrics}
We quantified task performance using two primary metrics:

\begin{enumerate}
    \item Accuracy: The proportion (\%) of correct selections of the lower-delay robot.
    \item Confidence calibration: AUROC2 computation measures how well participants' confidence ratings align with their actual performance. Algorithm~\ref{alg:auroc2} presents a detailed computation of AUROC2. \cite{fleming2014measure}
\end{enumerate}

These metrics allowed us to evaluate both the participants' ability to discriminate between the robots based on delay and their metacognitive awareness of their performance. By combining accuracy and confidence calibration measures, we aimed to gain comprehensive insights into human decision-making and self-assessment in this human-robot interaction task. This enables us to understand not just how well participants performed, but also how accurately they judged their own performance across varying levels of task difficulty.

\input{sections/experiment/2-1-staircase}
\RestyleAlgo{ruled}
\begin{algorithm}[t] 
\caption{Confidence Calibration (AUROC2)}\label{alg:auroc2}
\DontPrintSemicolon
\KwIn{
$correct$: vector of size $1 \times n_{trials}$, with $0$ for error and $1$ for correct trials\;
$conf$: vector of size $1 \times n_{trials}$, with confidence ratings from $1$ to $N_{ratings}$\;
$N_{ratings}$: number of available confidence levels
}
\KwOut{$auroc2$: type-2 area under the ROC curve}
Initialize $i \gets N_{ratings} + 1$\;
\For{$c \gets 1$ \KwTo $N_{ratings}$}{
    $H2[i-1] \gets \text{count}(conf = c \land correct) + 0.5$\;
    $FA2[i-1] \gets \text{count}(conf = c \land \lnot correct) + 0.5$\;
    $i \gets i - 1$\;
}
Normalize $H2 \gets H2 / \sum(H2)$\;
Normalize $FA2 \gets FA2 / \sum(FA2)$\;
Compute cumulative sums: $csum\_H2 \gets [0, \text{cumsum}(H2)]$, $csum\_FA2 \gets [0, \text{cumsum}(FA2)]$\;
Initialize $i \gets 1$\;
\For{$c \gets 1$ \KwTo $N_{ratings}$}{
    $k[i] \gets (csum\_H2[c+1] - csum\_FA2[c])^2 - (csum\_H2[c] - csum\_FA2[c+1])^2$\;
    $i \gets i + 1$\;
}
Compute $auroc2 \gets 0.5 + 0.25 \times \sum(k)$\;
\Return $auroc2$
\end{algorithm}

%% file: sections/results/results.tex
\section{RESULTS}
From the data obtained from 80 participants, we combine pairwise participant data to form virtual dyads~\cite{koriat2012two}.
We obtained 3160 virtual dyads $80 \choose 2$.
For each virtual dyad, we selected those trials where both the participants faced the same delay pair.
From these, we eliminated those trials where both participants agreed on their choice of robot, and further analysed those trials where the participants differed in their choice such that a joint decision was required. 


The pairing process (\ref{procedure}) was repeated for all delay-pair conditions and 4,148 virtual joint decision-making trials were generated from real data. Of these, 1,932 trials involved disagreements, requiring joint decisions for this dyad.

Within each dyad, the member with a higher percentage of correct responses was designated as high-performing (HP), while the other was designated as low-performing (LP). 
Additionally, three dummy participants were created: dummy high-confidence (D-HC), dummy low-confidence (D-LC), and dummy random (D-Random).
For each trial, the response of the participant who indicated higher confidence was selected as D-HC participant, and the other as the D-LC participant.
\textbf{D-HC corresponds to the Maximum Confidence Slating (MCS) approach of joint decision-making}.
D-Random represented a joint decision-maker who picks randomly between the choices made by both the participants in the dyad. 
Accuracy was then calculated for these five participants.

We then present our analysis formulated from the four backbone research questions. Formally, the independent variables are (1) individual confidence, (2) performance discrepancy, and (3) confidence calibration. The dependent variables include joint decision accuracy, accuracy gains from MCS, and the dyadic AUROC2 value measuring the dyad's confidence calibration.

\input{sections/results/subsections/intro}

\input{sections/results/subsections/mcs}

\input{sections/results/subsections/skill}

\input{sections/results/subsections/calibr}

\input{sections/results/subsections/goodbad}

\input{sections/results/subsections/dyad}

%% file: sections/results/subsections/intro.tex

%% file: sections/results/subsections/mcs.tex
\subsection{Impact of Confidence on Accuracy}
\par
\textit{\textbf{RQ1:} How does individual confidence affect the accuracy of joint decisions in dyadic settings, and what are the benefits of joint decision-making using MCS?}\vspace{2pt}
\par
\noindent

\input{sections/results/figs/overall}

%% file: sections/results/figs/overall.tex

\begin{figure}[t]
    \centering
    \includegraphics[width=0.9\columnwidth]{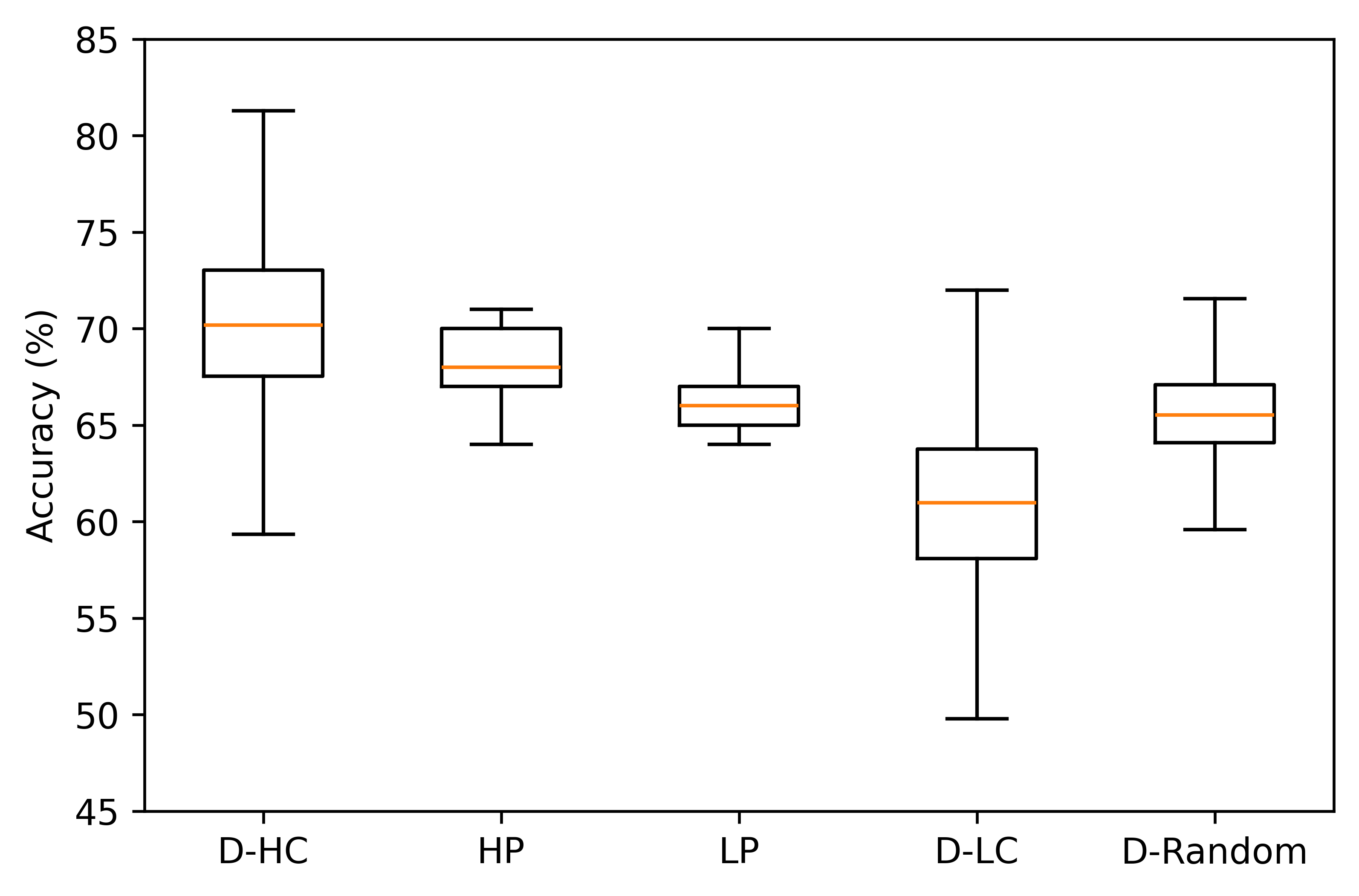}
    \caption{Accuracy in robot selection for high-confidence, higher-performing, low-confidence, lower-performing, and random participants from the paired dyads.}
    \label{fig:overall}
\end{figure}

Figure~\ref{fig:overall} illustrates the accuracy of robot selection across all five participant types, based on 3,160 virtual dyads. We assessed the performance of D-HC against other participants using the two-sample t-test. The results show that D-HC exhibited significantly higher accuracy compared to HP (Highest Performing individual), $(t(3158) = 8.36, p < 0.0001)$. This confirms that decisions made using the Maximum Confidence Slating (MCS) approach provide a significant advantage over those made by the highest-performing individual within the dyad. The accuracy of D-HC was also significantly greater than that of D-Random, $(t(3158) = 33.48, p < 0.0001)$.
Interestingly, D-Random performed better than D-LC, $(t(3158)$ $ = 33.48, p < 0.0001)$. This suggests that, in any  given scenario, a low-confidence response should not be favored. Additionally, we confirmed that D-HC had significantly higher accuracy than D-LC, $(t(3158) = 62.06, p < 0.0001)$, reinforcing that high-confidence decisions lead to better outcomes than low-confidence ones.

Through this analysis, we validated \textbf{H1: MCS is effective in robot teleoperation control selection task. Picking the higher confidence decisions within a dyad results in more accurate joint decisions than those made by the highest-performing individual alone. Conversely, low-confidence decisions yield accuracy levels comparable to random choices.}

%% file: sections/results/subsections/skill.tex
\subsection{Effect of Performance Discrepancy}
\par
\textit{\textbf{RQ2:} How do performance discrepancies between dyad members affect the accuracy gains from MCS-based joint decisions?} \vspace{2pt}
\par
\noindent
While MCS-based joint decision-making delivers a benefit overall, specific ways of pairing participants provides further insights.
In this section, we analyse the impact of pairing participants according to their performance level. We sorted the 3160 virtual dyads according to the \textbf{difference} in the performance of both individuals in each dyad, as measured using two factors: 
\begin{enumerate}
    \item Success rate $(\%)$ over the 100 trials. 
    \item Settling task difficulty level via the two-down one-up procedure.
\end{enumerate}

\input{sections/results/rssfigs/skillfigs}

%% file: sections/results/rssfigs/skillfigs.tex
\begin{figure}[t]
    \centering
    \includegraphics[width=0.9\columnwidth]{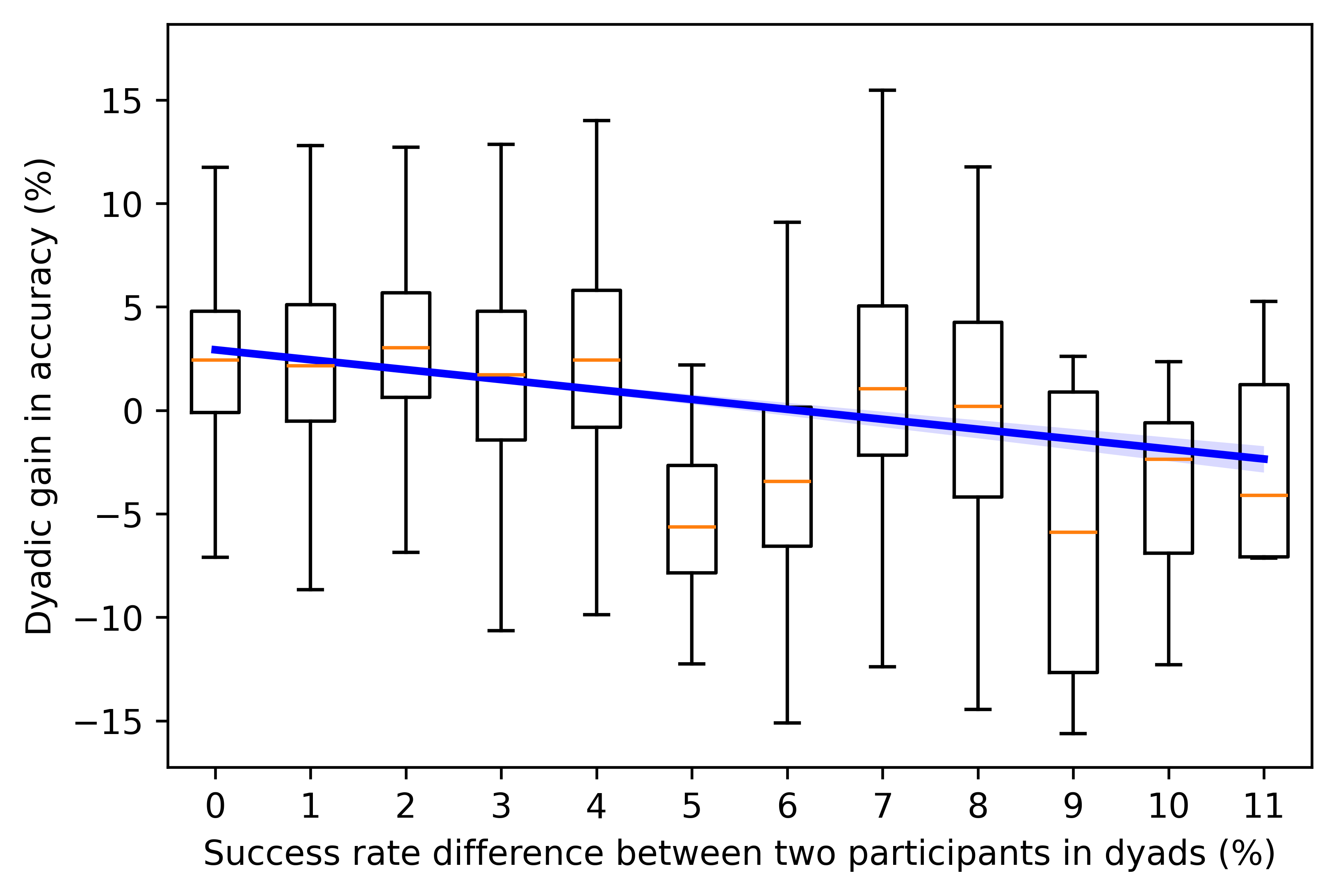}
    \caption{Variation in accuracy gain as participants of different success rates are paired together. Pairing similar performing participants leads to higher benefit.}
    \label{fig:succratediff-vs-accgain}
\end{figure}
\begin{figure}[t]
    \centering
    \includegraphics[width=0.85\columnwidth]{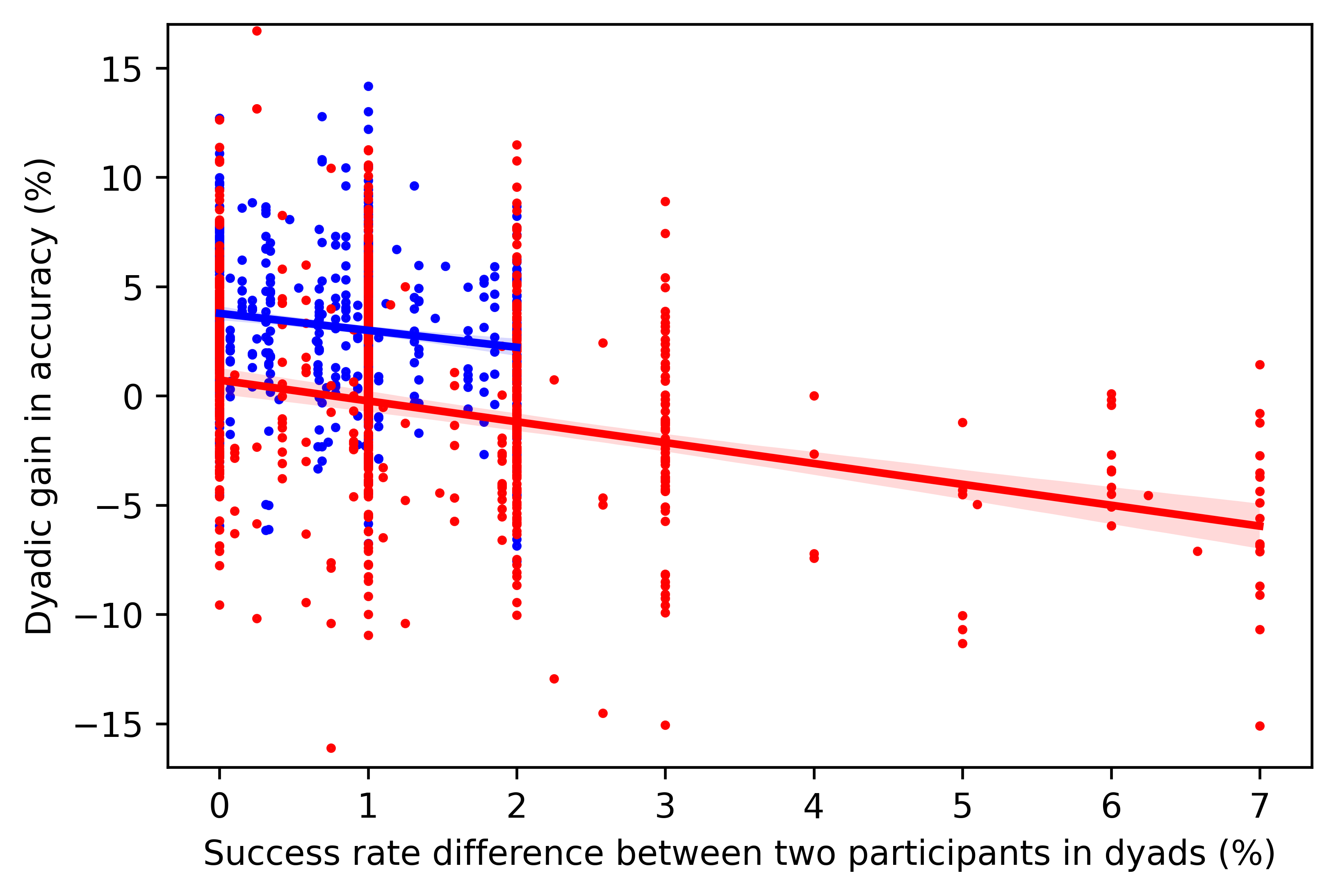}
    \caption{Variation in accuracy gain as participants of different success rates are paired together. Blue: both participants being below the mean skill level,
and red: both participants being above the mean skill level.}
    \label{fig:successrate-meandivided-accgain}
\end{figure}

\begin{figure*}[t]
\centering
\begin{subfigure}{\columnwidth}
  \centering
  \includegraphics[width=0.85\columnwidth]{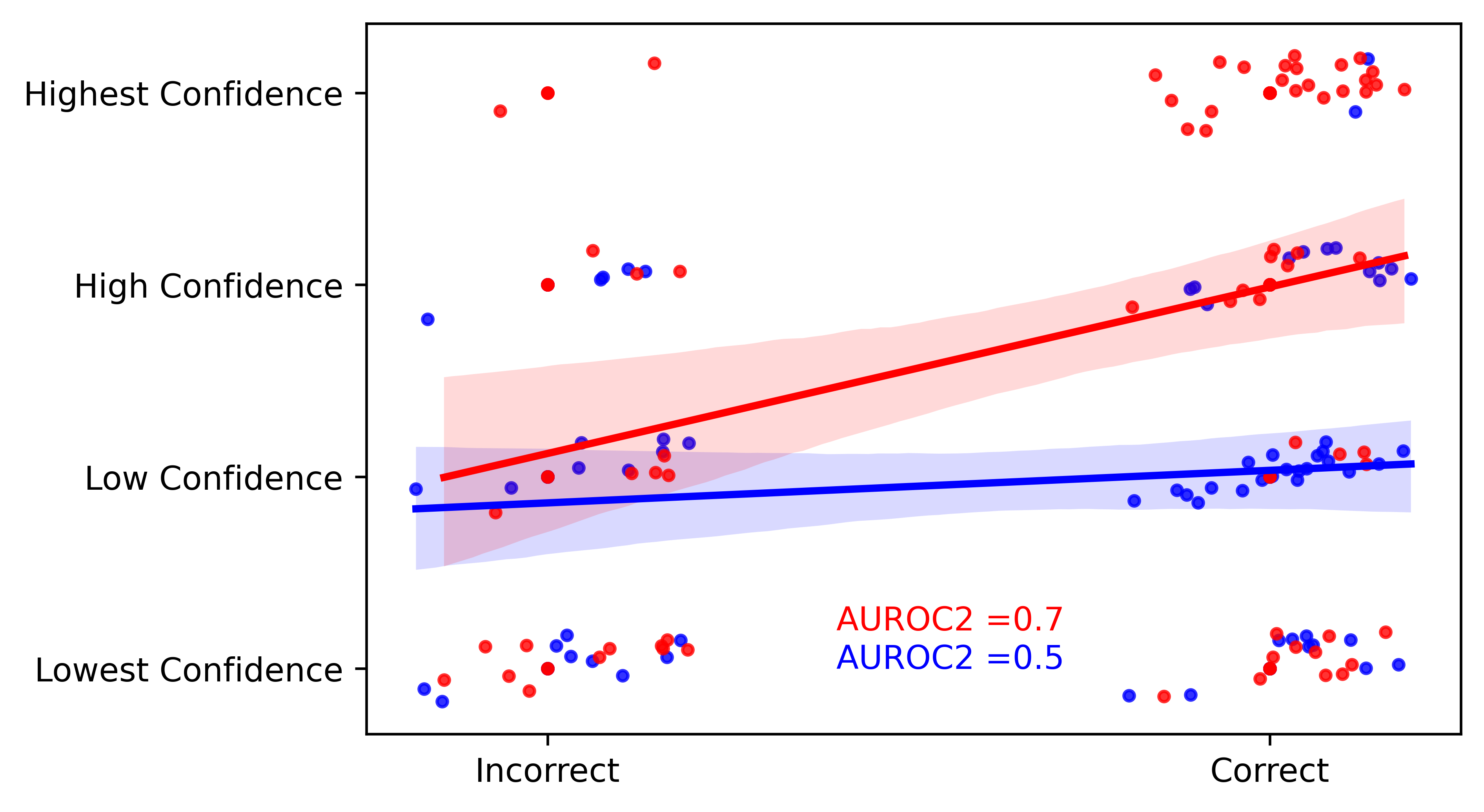}
  \caption{Correlation between correctness and confidence.}
  \label{fig:confcalib_good}
\end{subfigure}%
\begin{subfigure}{\columnwidth}
  \centering
  \includegraphics[width=0.7\columnwidth]{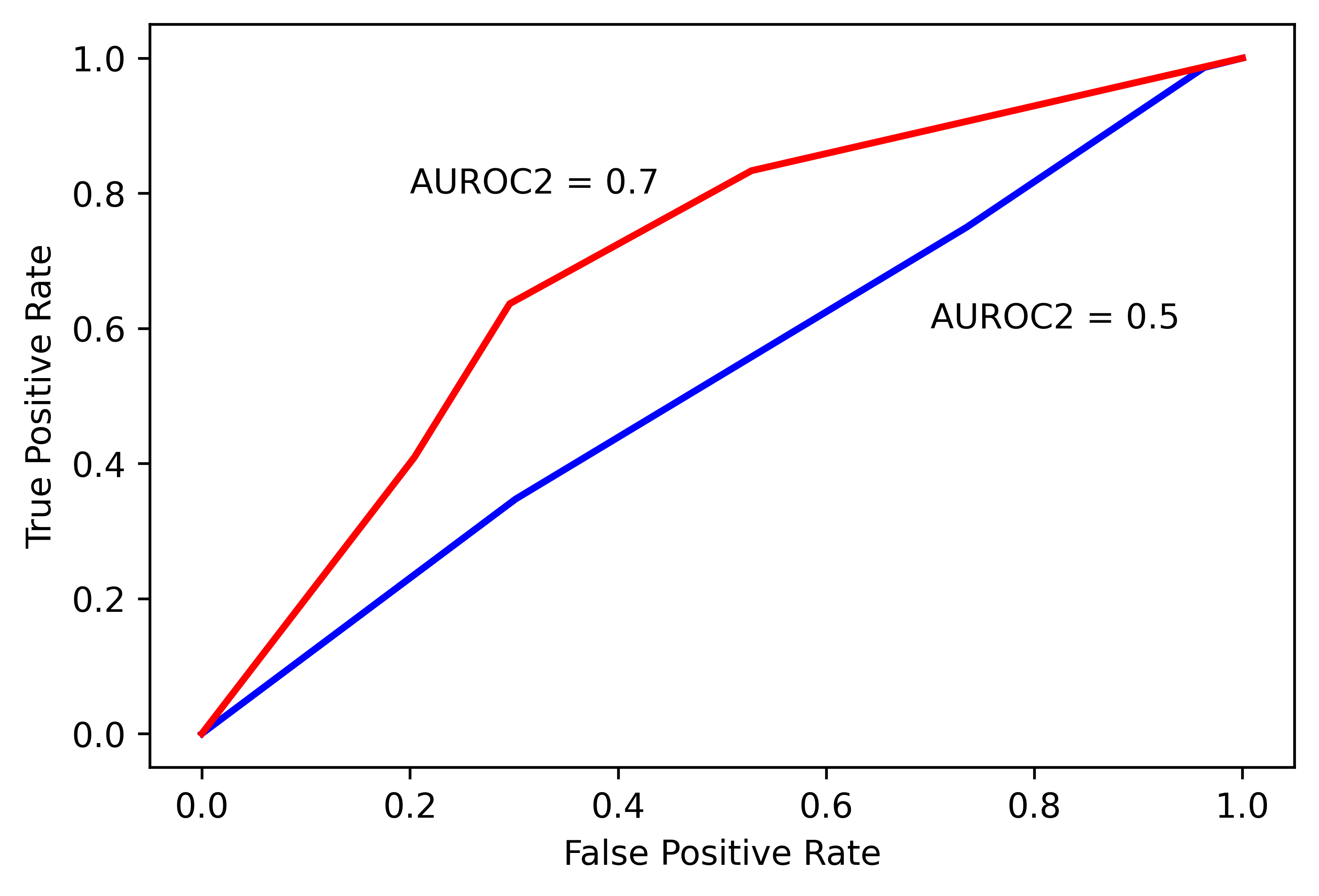}
  \caption{Corresponding AUROC2 curves.}
  \label{fig:auroc-curve-2pids}
\end{subfigure}
\caption{Confidence calibration of two different participants. \ref{fig:confcalib_good}: Plot of trial result vs. confidence rating. The line is the regression fit with 95\% confidence interval. \ref{fig:auroc-curve-2pids}: ROC curves for both participants. The participant represented by red has better calibrated confidence as compared to the participant represented by blue.}
\label{fig:confcalib}
\end{figure*}
\begin{figure}[h]
    \centering
    \includegraphics[width=0.8\columnwidth]{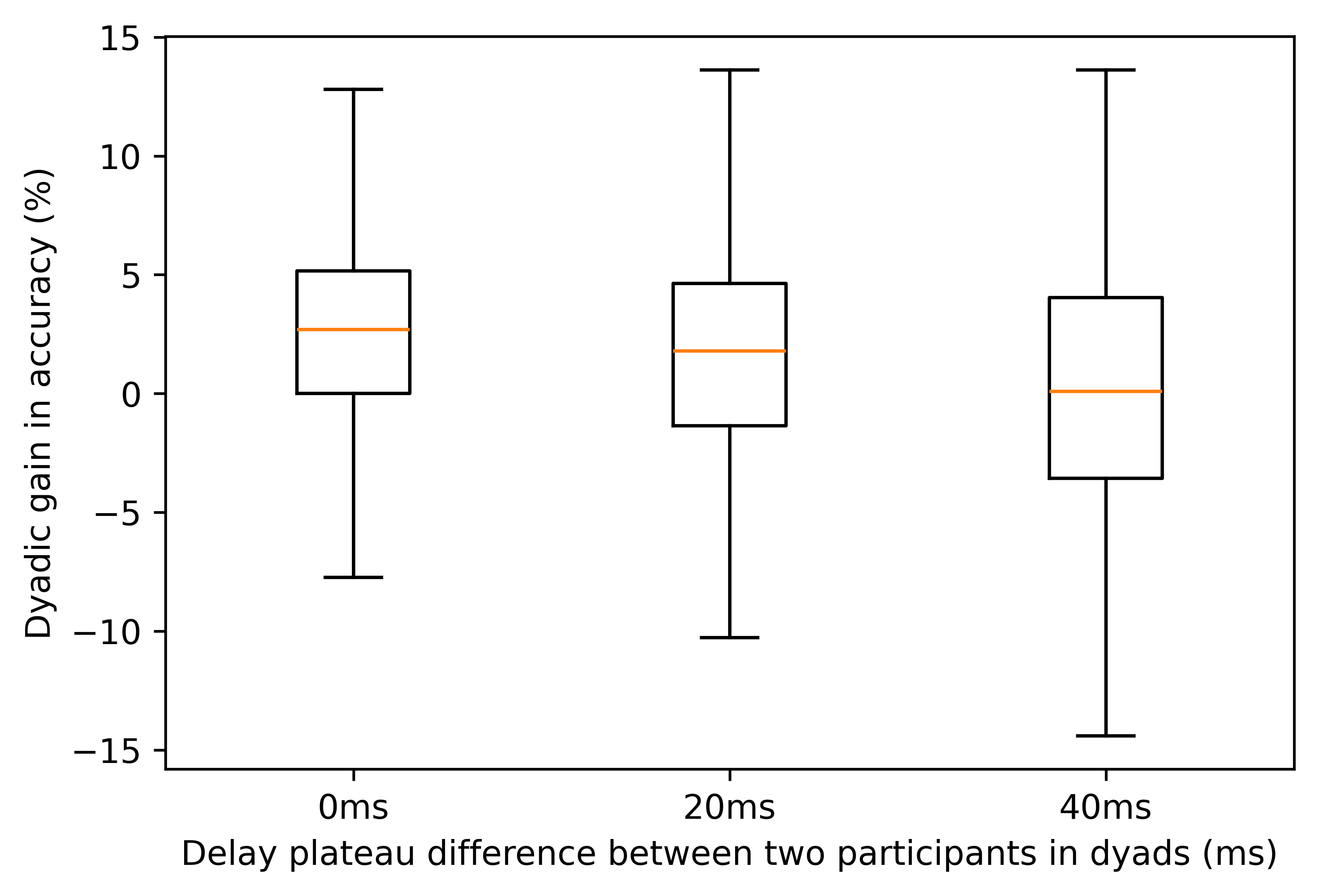}
    \caption{Variation in accuracy gain as participants of different performance levels are paired together. 
    Performance level is assessed by the difficulty of the task as established by delay plateau difference. Pairing similarly performing participants leads to higher accuracy gain.}
    \label{fig:taskdifficdiff-vs-accgain}
\end{figure}

For every virtual dyad, we extracted the corresponding \textit{accuracy gain}~\cite{Bang_Bahrami_2014}. Figure~\ref{fig:succratediff-vs-accgain} shows the impact of difference in performance level on the resulting accuracy gain of the dyad.
The difference in success percentage between the individuals in each virtual dyad ranged from 0 to 11\%.
As the dissimilarity in success rate increases, the resulting accuracy gain decreases $(\text{regression coefficient}: r = -0.48, p < 0.0001)$.

The accuracy gain corresponding to the dyads with no difference in success rate (i.e., identical skill level) was significantly higher than both dyads with a $5\%$ difference in success rate ($t(1008) = 7.1, p < 0.0001)$), as well as dyads with a $10\%$ difference in success rate ($(t(694) = 8.35, p < 0.0001)$).
Pairing participants with a success rate difference of more than 8\%, leads to a negative accuracy gain, i.e., MCS causes poorer joint performance as compared to the better performing individual.

For a more granular understanding of the impact of skill difference (similarly skilled can correspond to both being highly skilled or both being poorly skilled), we divided the participant data into two groups: above mean skill level (success percentage $>$ 67\%), and below mean skill level (success percentage $<$ 67\%).
Figure~\ref{fig:successrate-meandivided-accgain} shows the variation in accuracy gain with difference in performance of the two individuals for the two groups of participants.
While the trend is similar for both groups (\textcolor{blue}{blue}: both participants being below the mean, and \textcolor{red}{red}: both participants being above the mean), the absolute values of the accuracy gains show that when both participants are below mean performance the benefit provided by MCS-based joint decision is higher.
When both participants are above mean performance, the benefit is more observable for participants of similar performance level (lower success rate difference).


Figure~\ref{fig:taskdifficdiff-vs-accgain} shows the impact of pairing based on performance, measured by the task's settled difficulty level. Virtual dyads are grouped into three categories based on the delay difference faced by the two participants when task difficulty plateaus. A $0$ ms value on the x-axis indicates similar performance between participants, while a $40$ ms value indicates differing performance levels.
We see that the resulting accuracy gain is again higher for participants with similar performance level.
The accuracy gain is higher for the virtual dyad with similarly performing participants ($0$ ms) as compared to both $20$ ms $(t(5752) = 9.67, p < 0.0001)$ as well as $40$ ms $(t(3280) = 11.04, p < 0.0001)$.

This analysis supports \textbf{H2: Similar task performance leads to higher accuracy gains. Specifically, the accuracy gains from MCS are greater in dyads where the performance levels of the members are similar, compared to dyads with significant performance discrepancies. This is the first time it is observed that larger performance gaps between dyad members result in smaller accuracy gains.}

%% file: sections/results/subsections/calibr.tex
\subsection{Influence of Confidence Calibration}
\par
\textit{\textbf{RQ3:} How does the confidence calibration of participants influence the accuracy gains from MCS-based decisions?}
\par \vspace{2pt}
\noindent To find the answer to this, we paired participants according to confidence calibration and analysed the confidence calibration of participants to gain further insights into the performance of the MCS approach to joint decision-making.

\input{sections/results/figs/auroc-hist}

\input{sections/results/figs/confcalib}

\input{sections/results/rssfigs/pairbyaurocdiff}

%% file: sections/results/figs/auroc-hist.tex

%% file: sections/results/figs/confcalib.tex

\begin{figure*}[h]
\centering
\begin{subfigure}{0.65\columnwidth}
  \centering
  \includegraphics[width=\columnwidth]{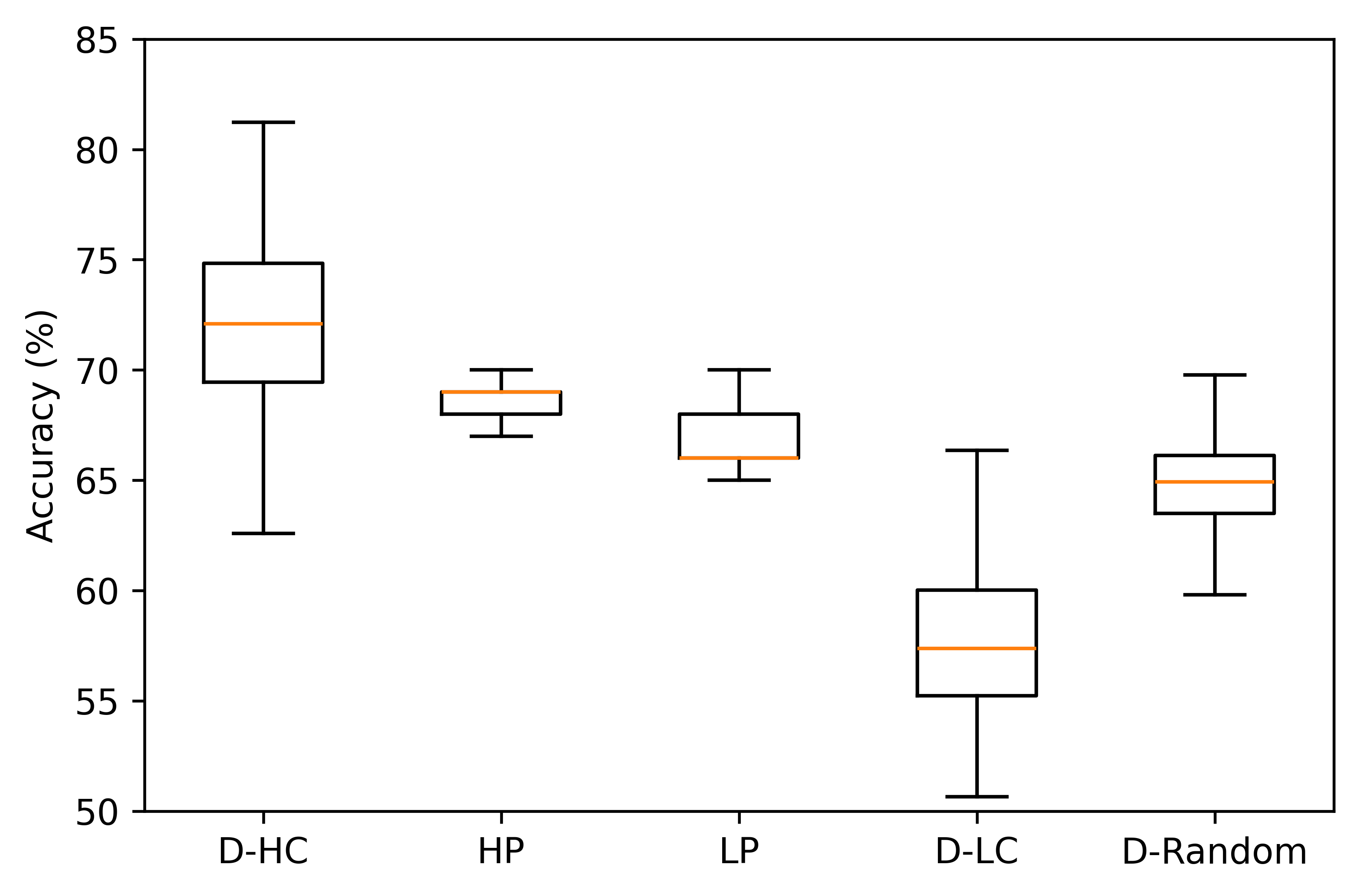}
  \caption{}
  \label{fig:good-good}
\end{subfigure}%
\begin{subfigure}{0.65\columnwidth}
  \centering
  \includegraphics[width=\columnwidth]{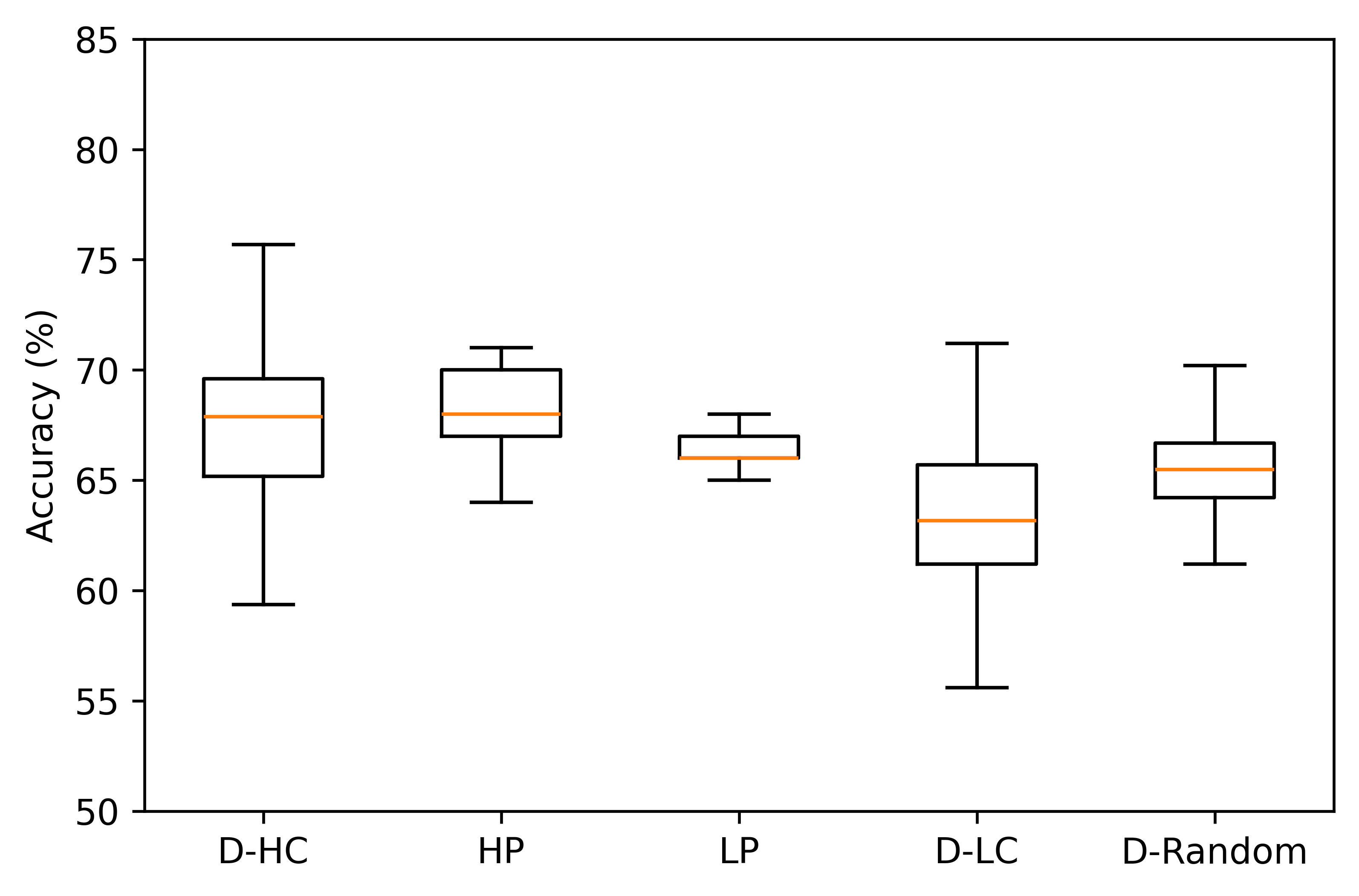}
  \caption{}
  \label{fig:bad-bad}
\end{subfigure}
\begin{subfigure}{0.65\columnwidth}
  \centering
  \includegraphics[width=\columnwidth]{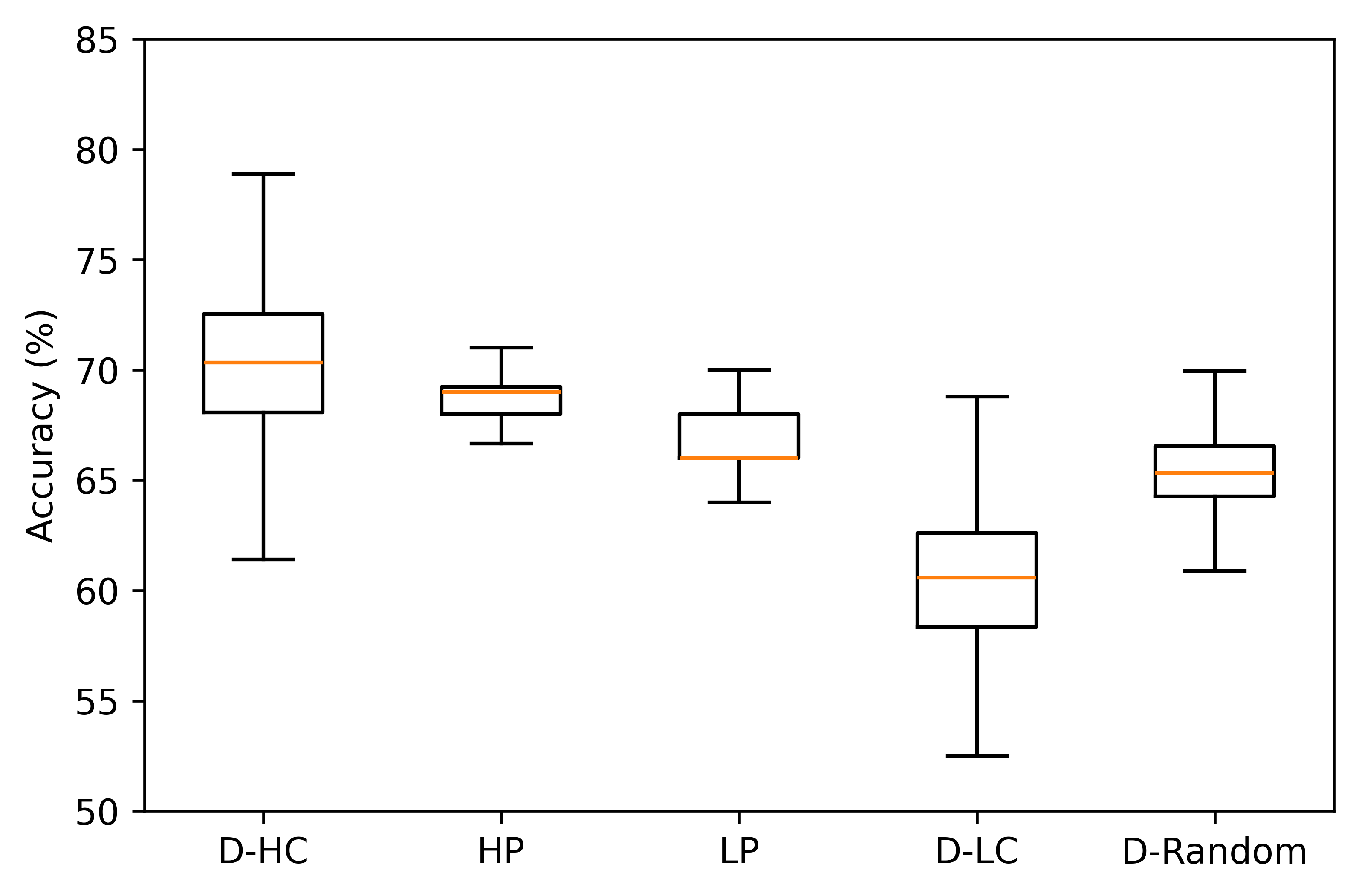}
  \caption{}
  \label{fig:good-bad}
\end{subfigure}
\caption{Choice accuracy (\%) of different combinations of participants according to their confidence calibration: \ref{fig:good-good}: both well calibrated (288 pairs), \ref{fig:bad-bad}: both poorly calibrated (242 pairs), and \ref{fig:good-bad}: well calibrated with poorly calibrated (264 pairs).}
\label{fig:goodbad}
\end{figure*}

\begin{figure}[h]
    \centering
    \includegraphics[width=0.85\columnwidth]{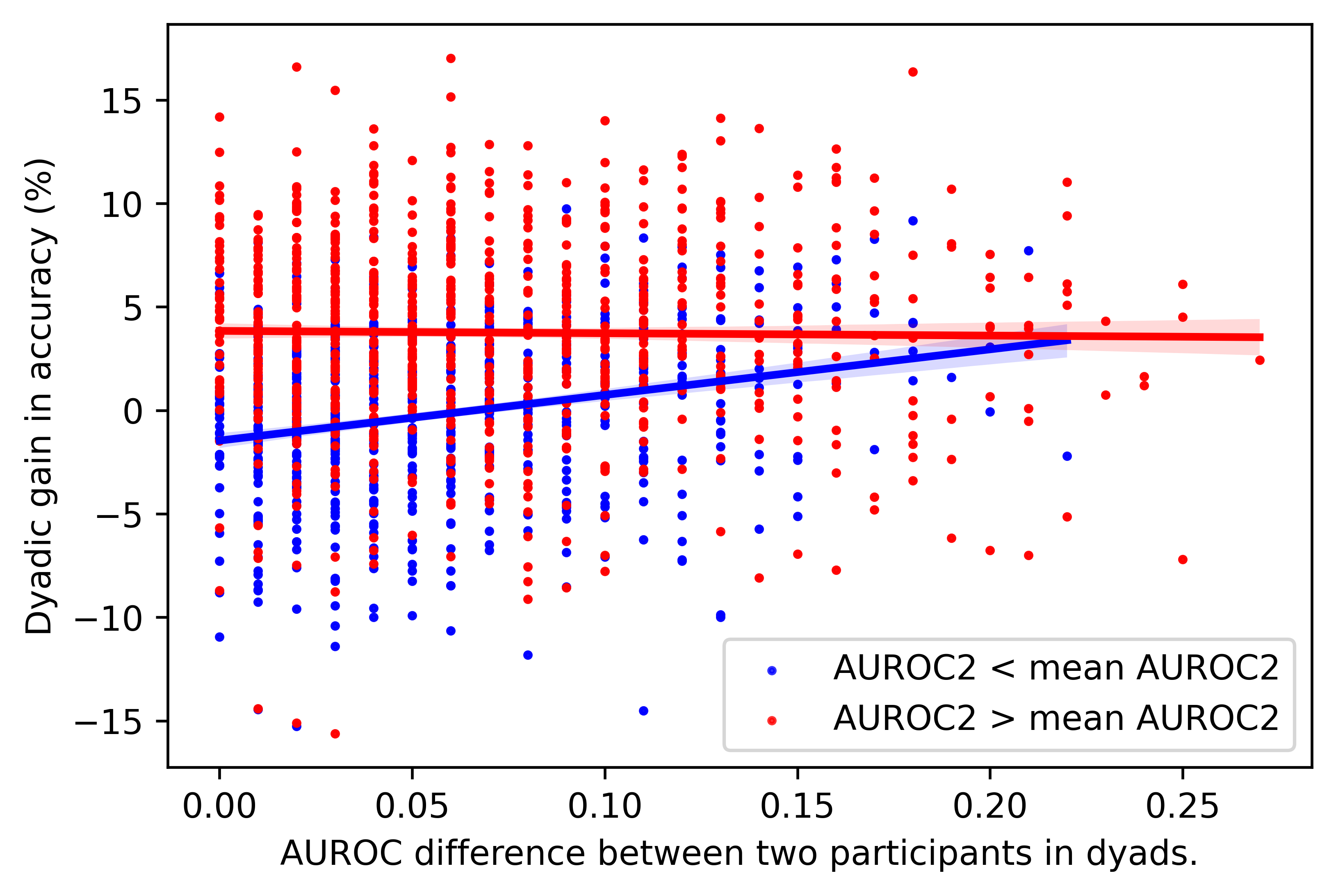}
    \caption{Variation in accuracy gain as participants of different confidence calibration are paired together. For well-calibrated individuals (red), pairing participants of similar confidence calibration leads to higher benefit. Pairing similarly calibrated participants who are poorly calibrated (blue) does not yield a benefit.}
    \label{fig:aurocdiff-vs-accgain}
\end{figure}

Figure~\ref{fig:confcalib} illustrates the variation of confidence rating with the correctness of response for two different participants at either end of the spectrum of AUROC2 values.
The (\textcolor{red}{red}) participant has better calibration (AUROC2 = 0.7) than the (\textcolor{blue}{blue}) participant (AUROC2 = 0.5).
Figure~\ref{fig:confcalib_good} shows a plot of trial result versus associated confidence value and Figure~\ref{fig:auroc-curve-2pids} shows the corresponding ROC curves for both the participants.

%% file: sections/results/rssfigs/pairbyaurocdiff.tex
To analyze the impact of confidence calibration similarity on accuracy gain, participants were grouped based on their AUROC2 values: above the mean (AUROC2 $>$ 0.6) and below the mean (AUROC2 $<$ 0.6).We then paired participants within the same group and analysed the impact on accuracy gain. Figure~\ref{fig:aurocdiff-vs-accgain} shows that for participants with AUROC2 above the mean, MCS-based joint decisions with similarly calibrated pairs improved accuracy. However, for those below the mean, pairing similar participants led to negative accuracy gain, with MCS performing worse than the better individual.

%% file: sections/results/subsections/goodbad.tex
\input{sections/results/figs/goodbad}

%% file: sections/results/figs/goodbad.tex
We further investigated the impact of pairing participants according to confidence calibration by selecting the top participants from the group with AUROC2 values above the mean (AUROC2 $>$ 0.65: well-calibrated, 25 participants), and bottom participants from the group with AUROC2 values below the mean (AUROC2 $<$ 0.55: poorly-calibrated, 26 participants).
We formed virtual dyads by pairing participants according to their confidence calibration: both well-calibrated participants (288 virtual dyads); both poorly-calibrated participants (242 virtual dyads); and well-calibrated participant paired with poorly-calibrated participant (264 virtual dyads).



Figure~\ref{fig:goodbad} presents the performance accuracy for the five types of participants (HP, LP, D-HC, D-LC, and D-Random) in the virtual dyads. Figure~\ref{fig:good-good} shows the accuracy when both individuals in the dyad are well-calibrated in terms of confidence. The performance of D-HC was significantly better than HP, $(t(574) = 1.39, p < 0.0001)$. This demonstrates that pairing two well-calibrated individuals results in an increased choice's accuracy compared to the higher performance individual. However, this is not the case when both individuals are poorly calibrated. Figure~\ref{fig:bad-bad} illustrates the accuracy for dyads with two poorly calibrated individuals, where HP outperformed D-HC, $(t(482) = 6.08, p < 0.0001)$. This suggests that in such cases, it is preferable to rely on the more performant individual’s choice. Figure~\ref{fig:good-bad} presents the results when a well-calibrated participant is paired with a poorly calibrated one. In this case, D-HC outperformed HP, $(t(526) = 10.86, p < 0.0001)$, indicating that the joint decision was more accurate when at least one individual had good confidence calibration.

Through this analysis, we confirmed \textbf{H3: Participants with above-average confidence calibration achieve stable accuracy gains from MCS, regardless of the calibration differences between dyad members. Conversely, participants with below-average calibration show improved accuracy when paired with individuals who have diverse calibration levels.}


%% file: sections/results/subsections/dyad.tex
\subsection{Dyadic Confidence-Calibration Correlation}
\par
\textit{\textbf{RQ4:} What is the relationship between dyadic confidence calibration and the accuracy of MCS-based joint decisions?} \vspace{2pt}
\par




\input{sections/results/figs/dyad-confcalib}

%% file: sections/results/figs/dyad-confcalib.tex
\begin{figure}[t]
    \centering
    \includegraphics[width=0.9\columnwidth]{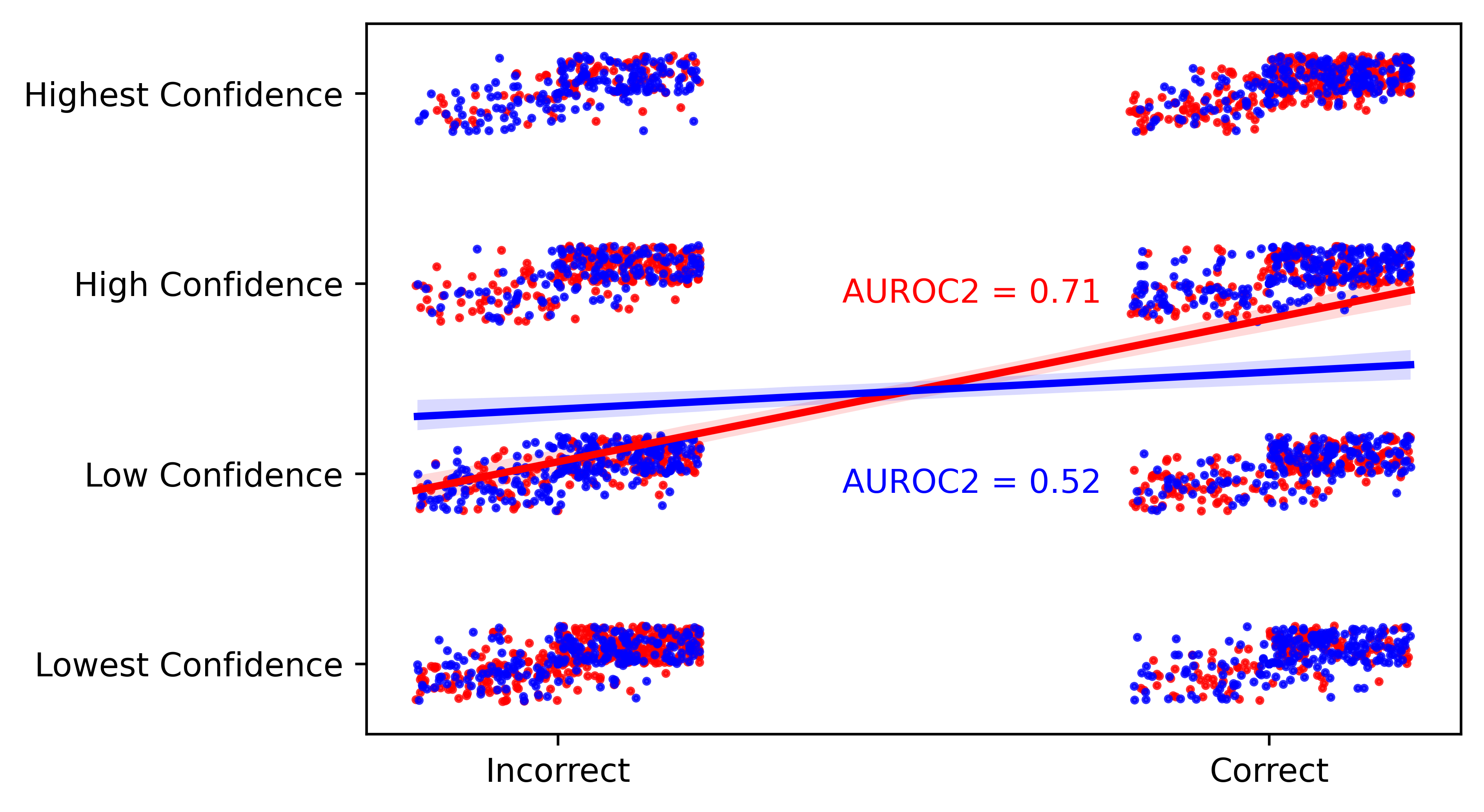}
    \caption{Trial results versus confidence ratings for example virtual dyads. Red: a dyad formed by pairing well-calibrated participants. Blue: a dyad formed by pairing poorly calibrated participants.}
    \label{fig:dyad-conf-calibration}
\end{figure}

\noindent We further investigated the behavior of resulting dyads formed from pairing individuals together. 
We computed dyadic confidence calibration using Koriat’s method~\cite{koriat2012two}, applying the AUROC2 computation (Algorithm.~\ref{alg:auroc2} ) to the concatenated decision-confidence pairs of both participants. For instance, Participant 1 completed 20 trials and Participant 2 completed 15 trials under the same delay condition (50 ms for Robot 1, 70 ms for Robot 2). Of the combined 35 trials, 6 with identical robot choices were excluded, leaving 29 trials for dyadic confidence calibration computation. Figure~\ref{fig:dyad-conf-calibration} shows the trial result vs. confidence rating for an example participant dyad, along with the associated AUROC2 value.



\begin{figure}[t]
    \centering
    \includegraphics[width=0.8\columnwidth]{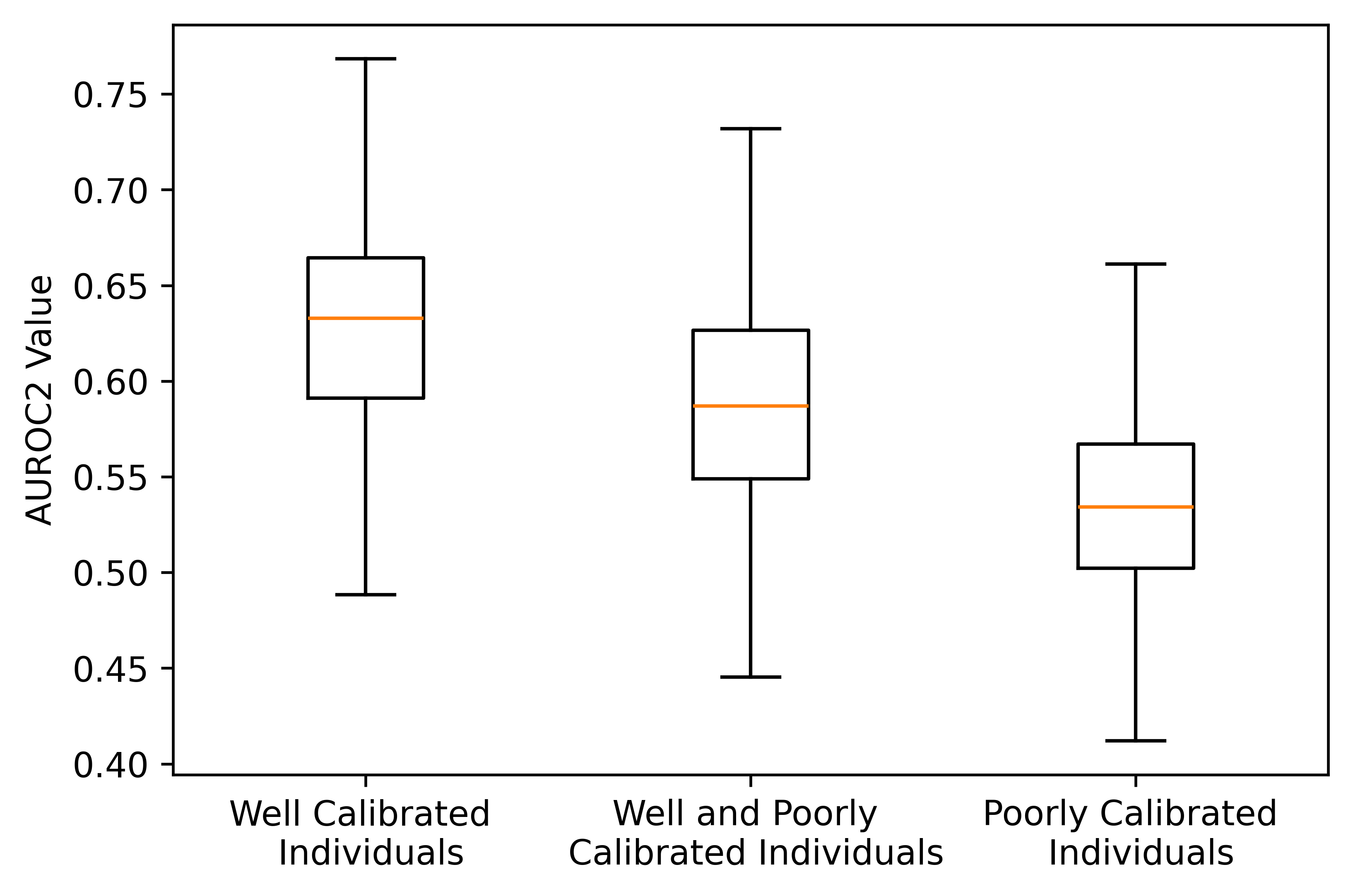}
    \caption{Variation in dyadic confidence calibration based on how participants are paired according to their individual confidence levels.}
    \label{fig:dyad-auroc}
\end{figure}

The calibration of individuals in a dyad affects the resulting dyad's confidence calibration. Figure~\ref{fig:dyad-auroc} shows dyadic confidence calibration based on individual calibrations. Statistical analysis reveals that dyads of two well-calibrated individuals have higher AUROC2 than those with both poorly calibrated individuals $(t(1012) = 28.74, p < 0.001)$ or mixed-calibration dyads $(t(1078) = 13.28, p < 0.001)$.

\begin{figure}[t]
    \centering
    \includegraphics[width=0.8\columnwidth]{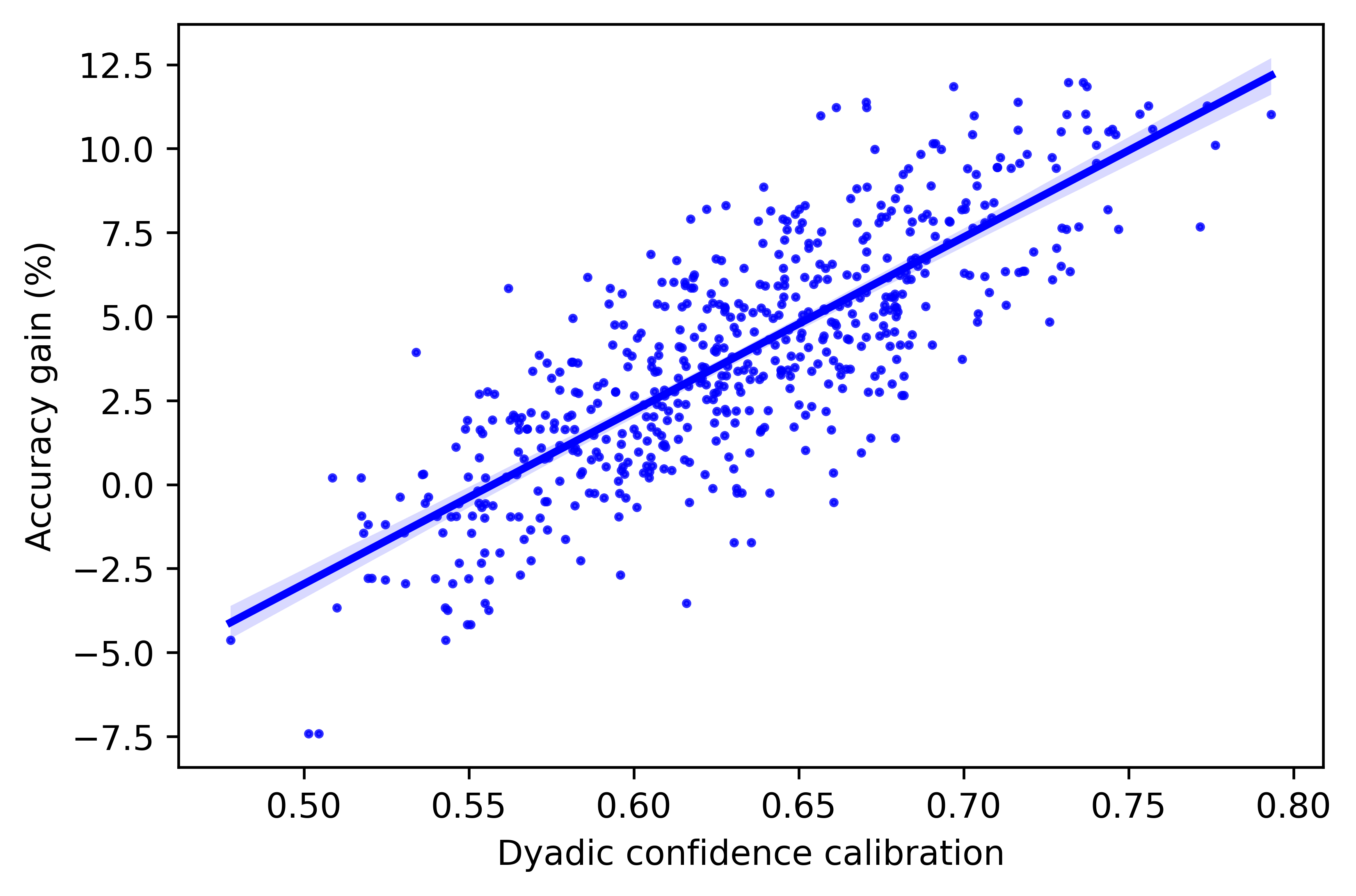}
    \caption{Dyadic confidence calibration versus accuracy gain. Higher dyadic confidence calibration leads to an increased accuracy gain.}
    \label{fig:gain}
\end{figure}

Finally, our findings highlight the significant impact of dyadic AUROC2 on joint decision-making performance. Figure~\ref{fig:gain} shows a strong correlation between accuracy gain and dyadic confidence calibration, with improved calibration leading to higher joint decision accuracy. This underscored the importance of well-calibrated confidence within the dyad for the effectiveness of the Maximum Confidence Slating (MCS) approach.

This analysis supports our fourth hypothesis \textbf{H4: Dyads with higher overall confidence calibration demonstrate better decision accuracy when using MCS-based joint decision, compared to dyads with lower calibration. This finding reinforces the value of pairing individuals with well-calibrated confidence for optimal team performance.}

%% file: sections/discussion/discussion.tex
\section{DISCUSSION}

\subsection*{Summary:}In this research, we investigated human-human dyad joint decision-making in a robot teleoperation task, focusing on how maximum confidence slating (MCS) choice selection impacts decision accuracy. To our knowledge, this is the first study applying MCS-based joint decision-making in a dynamic, spatiotemporal task involving active robot control. Our results showed that the accuracy of dyad joint decisions was significantly higher than that of the more skilled individual in the pair. Our findings emphasise the importance of skill similarity and confidence calibration in achieving better outcomes for human collaboration in robotic tasks, establishing a foundational understanding of MCS's role in the human-robot interaction domain. The effectiveness of MCS in this context highlights its potential for improving joint decision-making in human-IDS (Intelligent Decision Support) system dyads. This initial work also reveals the potential of MCS for real-world applications requiring critical, time-sensitive decisions. By leveraging confidence as a low-cost metric, MCS combines two operators' confidence levels to improve task efficiency and decision accuracy.


\subsection*{Limitations and Research Directions:}
The scope of our study was constrained to a specific robot teleoperation and controller selection task, where the primary decision involved selecting between two robots with different control delays. Our paper looked into virtual dyads scenarios, and in real joint decision-making between two people social effects like the perceived competence of oneself and of the other, differences in status can play a big role. Future research can expand to encompass a broader range of tasks and decision-making scenarios. 

One key area is extending our study to human-AI dyads. We plan to develop an Intelligent Decision Support (IDS) and AI systems tailored to the robot teleoperation task. By comparing findings from human-AI dyads with those from human-human dyads, we can gain valuable insights into how human confidence calibration influences decision-making when interacting with AI systems in robotic scenarios. 
Through these research directions, we hope to contribute to the integration of AI systems and development of more effective collaborative Human-Robot Interaction (HRI) settings.

\section{Acknowledgement}
This work was supported by the EPSRC Programme Grant ‘From Sensing to Collaboration’ (EP/V000748/1). Raunak Bhattacharyya acknowledges support from the Yardi School of AI publication grant.


